\documentclass[aps,prl,twocolumn,amsmath,amssymb,footinbib,showpacs,longbibliography,superscriptaddress]{revtex4-1}

\newcommand{\Jnature}{Nature (London)}

\newcommand{\Jscience}{Science}

\newcommand{\Jprl}{Phys. Rev. Lett.}

\newcommand{\Jpra}{Phys. Rev. A}
\newcommand{\Jprb}{Phys. Rev. B}

\newcommand{\Jpre}{Phys. Rev. E}
\newcommand{\Jrmp}{Rev. Mod. Phys.}

\newcommand{\Jepl}{Europhys. Lett.}

\newcommand{\JphyslettA}{J. Phys. Lett. A}

\newcommand{\Jprocroysoc}{Proc. Roy. Soc. A: Math. Phys. Eng. Sci.}

\newcommand{\JphysicaB}{Physica B}

\newcommand{\JAnnualRevCondMat}{Annual Rev. Cond. Mat. Phys.}

\usepackage{times}

\usepackage[english]{babel}
\usepackage{latexsym}
\usepackage{graphics}
\usepackage{subfigure}
\usepackage{epsfig}
\usepackage{color}
\usepackage{hyperref}
\usepackage{braket} 
\usepackage[T1]{fontenc}
\usepackage[latin9]{inputenc}
\usepackage{amstext}
\usepackage{amssymb}
\usepackage{graphicx}
\usepackage{esint}
\usepackage{braket}
\usepackage{babel}
\usepackage{amsmath}
\usepackage{verbatim}
\usepackage{ulem}

\hypersetup{
colorlinks=true,
citecolor=blue,
linkcolor=red,
urlcolor=black
}


\newcommand{\gOneD}{g_\textrm{\tiny 1D}}

\newcommand{\aOneD}{a_\textrm{\tiny 1D}}

\newcommand{\kB}{k_\textrm{\tiny B}}

\newcommand{\ErOne}{E_\textrm{\tiny r1}}
\newcommand{\Eri}{E_\textrm{\tiny ri}}
\newcommand{\fs}{f_\textrm{\tiny s}}

\newcommand{\lettersection}[1]{\paragraph*{#1.---}}

\begin{document}

\title{Mott transition for a Lieb-Liniger gas in a shallow quasiperiodic potential: Delocalization induced by disorder
}

\author{Hepeng Yao}\email{Hepeng.Yao@unige.ch}
\affiliation{DQMP, University of Geneva, 24 Quai Ernest-Ansermet, CH-1211 Geneva, Switzerland}
\author{Luca Tanzi}
\affiliation{Istituto Nazionale di Ottica, CNR-INO, Via Moruzzi 1, 56124 Pisa, Italy}
\affiliation{European Laboratory for Non-Linear Spectroscopy,
Universit\`{a} degli Studi di Firenze, Via N. Carrara 1, 50019 Sesto Fiorentino, Italy}
\author{Laurent Sanchez-Palencia}
\affiliation{CPHT, CNRS, Ecole Polytechnique, IP Paris, F-91128 Palaiseau, France}
\author{Thierry Giamarchi}
\affiliation{DQMP, University of Geneva, 24 Quai Ernest-Ansermet, CH-1211 Geneva, Switzerland}
\author{Giovanni Modugno}
\affiliation{Istituto Nazionale di Ottica, CNR-INO, Via Moruzzi 1, 56124 Pisa, Italy}
\affiliation{European Laboratory for Non-Linear Spectroscopy,
Universit\`{a} degli Studi di Firenze, Via N. Carrara 1, 50019 Sesto Fiorentino, Italy}
\affiliation{Dipartimento di Fisica e Astronomia, Universit\`{a} degli Studi di Firenze,
Via G. Sansone 1, 50019 Sesto Fiorentino, Italy}
\author{Chiara D'Errico}\email{chiara.derrico@cnr.it}
\affiliation{Istituto per la Protezione Sostenibile delle Piante, CNR-IPSP, Strada delle Cacce 73, 10135 Torino, Italy}
\date{\today}

\begin{abstract}
Disorder or quasi-disorder is known to favor the localization in many-body Bose systems. Here in contrast, we demonstrate an anomalous delocalization effect induced by incommensurability in quasiperiodic lattices. Loading ultracold atoms in two shallow periodic lattices with equal amplitude and either equal or incommensurate spatial periods, we show the onset of a Mott transition not only in the periodic case but also in the quasiperiodic case. Upon increase of the incommensurate component of the potential we find that the Mott insulator turns into a delocalized superfluid. Our experimental results agree with quantum Monte Carlo calculations, showing anomalous delocalization induced by the interplay between the commensuration and interaction.

\end{abstract}

\maketitle

The Mott insulator (MI) is one of the most remarkable paradigmatic phases in strongly correlated quantum materials~\cite{Mott1937,Mott1949,Mott1956}.
It appears in condensed-matter systems when correlation effects associated with the strong electron-electron repulsion drive a metal-insulator phase transition~\cite{Imada1998}. The MI characterizes a broad class of materials~\cite{Lee2006,Furukawa2015,Cao2018,Tang2020,Regan2020,Shimazaki2020}, and is related to exotic quantum phenomena such as high-critical temperature superconductivity~\cite{Dagotto1994}, fractional quantum Hall effect~\cite{Burkov2010,Kuno2017}, and topological phase transitions~\cite{Sen2020}. MIs also appear in bosonic lattice models due to the competition between tunneling and repulsive interactions~\cite{Greiner2002}.
Experiments with ultracold atoms in optical lattices allow for in-depth investigation of many-body physics in a broad range of models~\cite{Bloch2008,Esslinger2010,Gross2017} and proved instrumental for direct observation and characterization of Mott phases for both Bose~\cite{Greiner2002,Gerbier2005,Gerbier2005PRA} and Fermi~\cite{Jorders2008,Schneider2008} systems,
first in three dimensions and later also in lower dimensional systems~\cite{Stoferle2004,Spielman2007,Spielman2008,Greif2013,Greif2015,Cocchi2016,Drewew2016}.
Remarkably, for one dimensional (1D) bosonic systems with sufficiently strong repulsive interactions, a purely periodic potential with arbitrary small amplitude can stabilize a Mott phase~\cite{haldane1980,haldane1981,giamarchi1997,giamarchi_book_1d,buchler2003}, as confirmed experimentally in Refs.~\cite{haller2010,Boeris2016}. 

Recently, quasiperiodic systems realized by two periodic lattices with incommensurate spatial periods have attracted a lot of attention. Incommensuration induces intriguing quantum phenomena such as Anderson localization~\cite{aubry1980,fallani2007,DErrico2014},
Bose glass (BG)~\cite{giamarchi1987,giamarchi1988,fisher1989}, and fractional MIs~\cite{Roux2008,Sanchez-Palencia2020}.
The phase diagrams of interacting bosons in such systems has been extensively studied theoretically both in one~\cite{Roux2008,roscilde2008,Sanchez-Palencia2020} and two~\cite{gautier-2Dquasicrystal-2021,zhu-2dquasicrystal-2022} dimensions and recent experiments have reported relevant measurements for 1D tight-binding models~\cite{DErrico2014,Gori2016,DErrico2021} and 2D quasicrystals~\cite{sbroscia-2dquasicrystal-2020,yu-2dquasicrystal-2023}.

\begin{figure}[b!] 
          \centering \includegraphics[width=1\columnwidth]{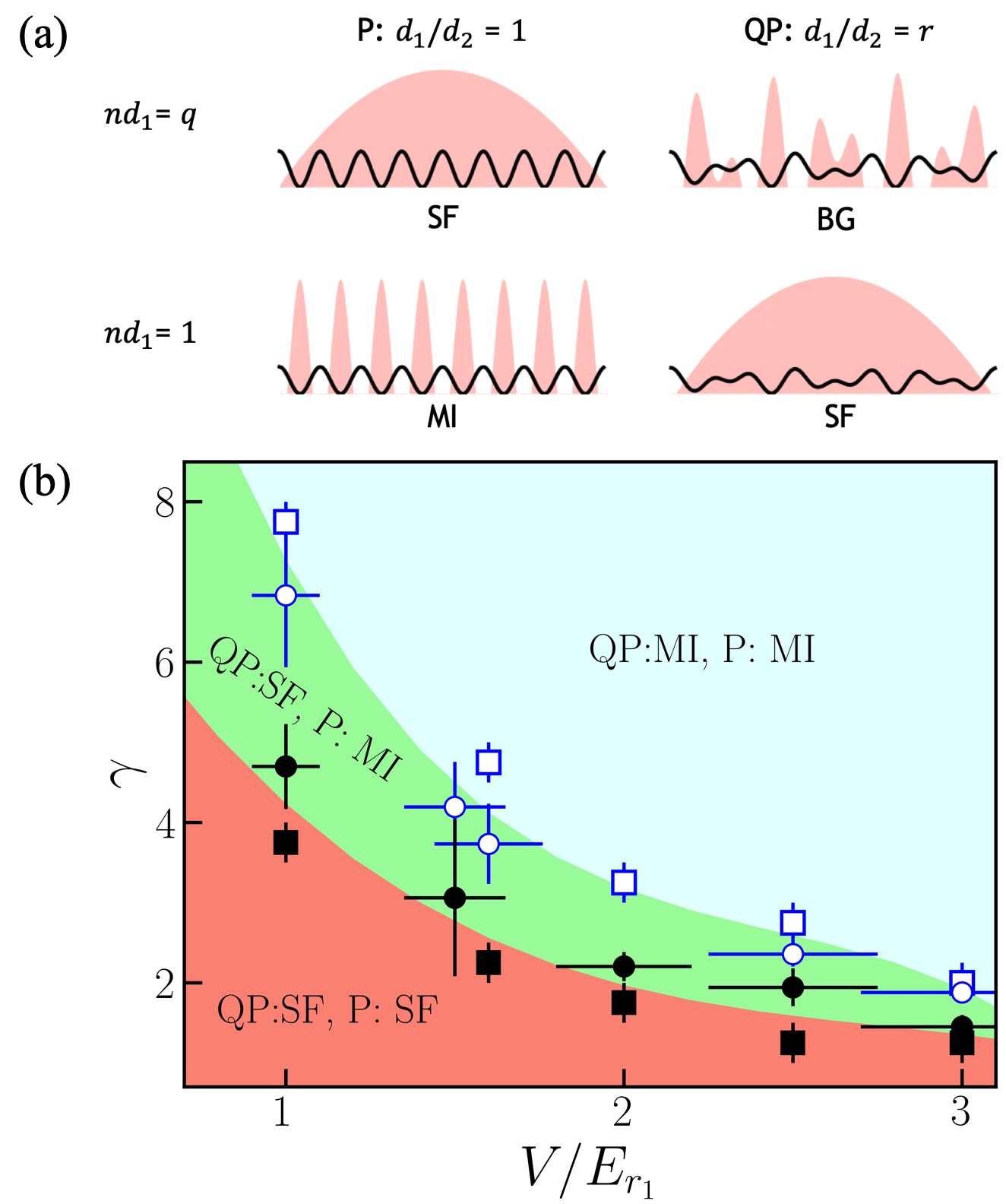}
           \caption{\label{fig:main} 
           Quantum phase transitions in periodic and quasiperiodic systems.
           (a)~Sketch of a strongly-interacting Bose gas for different commensuration: $nd_1=q$ (top) and $nd_1=1$ (bottom), for periodic (P; left) and quasiperiodic (QP; right) systems being $d_1$ the lattice period with larger spacing.
           (b)~Phase diagram for the MI to superfluid (SF) transition versus interaction strength and potential amplitude for unit particle filling $nd_1=1$. The transition lines are drawn by fits of the experimental (disk and circles) and QMC (squares) data points, for both periodic (solid markers) and quasiperiodic (hollow markers) systems.
           }
\end{figure}

Quasiperiodic lattices offer the possibility of studying the open problem of localization vs delocalization in presence of incommensuration. Bosons in a 1D periodic lattice with incommensurate filling always exhibits an extended superfluid phase at zero temperature.
Changing from a periodic to a quasiperiodic lattice, although keeping the total amplitude unchanged, the system will tend to localize and form a BG phase, as illustrated in the first row of Fig.~\ref{fig:main}(a). This suggests that quasiperiodicity favors localization.
However, when the number of particles is commensurate with the total number of the lattice sites,  the situation is completely opposite.
As illustrated in the second row of Fig.~\ref{fig:main}(a), while the periodic case favours a localized MI phase,
the quasiperiodic case will favour the delocalized superfluid phase.
While the Mott transition in periodic system has been assessed~\cite{haller2010,Boeris2016}, the quasiperiodic case, especially the delocalizing effect induced by incommensuration, is still a conjecture.

In this work, we experimentally observe the interplay of commensuration and localization in a strongly-correlated Bose gas using ultracold atoms in an optical lattice with switchable commensurability. We use two shallow periodic lattices with equal amplitude and either equal or incommensurate spatial periods. Using transport and excitation measurements, we show the onset of a Mott transition in the quasiperiodic case, at a different lattice amplitude than the one already observed in the periodic case~\cite{haller2010,Boeris2016}. The experiment is in good agreement with quantum Monte Carlo (QMC) calculations.
Comparing the phase diagrams in the periodic and quasiperiodic cases, both
theory and experiment show anomalous delocalization arising from incommensurability, see Fig.~\ref{fig:main}(b).

\lettersection{System and quantum phase diagram}
We consider a low temperature 1D gas of interacting bosons (Lieb-Liniger gas) in the presence of an external lattice potential. Its Hamiltonian writes
\begin{equation}\label{eq:Hamiltonian}
\mathcal{H} =\sum_{1 \leq j \leq N} \Big[-\frac{\hbar^2}{2m}\frac{\partial^2}{\partial z_j^2}+V(z_j)\Big]+\gOneD\sum_{j<\ell}\delta(z_j-z_\ell),
\end{equation}
with $m$ the atomic mass, $z_j$ the position of particle $j$, and $\gOneD=-2\hbar^2/m\aOneD$ the coupling constant, with $\aOneD$ the 1D scattering length.
In the experiment, 1D tubes are created by strong transverse confinement, and $\aOneD$ is related to the 3D scattering length $a$ and to the transverse oscillator length $\ell_{\perp}=\sqrt{\hbar/m\omega_{\perp}}$ via $\aOneD=\ell_{\perp}^2(1-1.03 a/\ell_{\perp})/a$~\cite{Olshanii1998,Cazalilla_2004}.
The quasiperiodic potential writes
\begin{equation}\label{eq:QPpotential}
V(z) = \frac{V}{2} \big[ \sin^2 \left(k_1 z\right) + \sin^2 \left(k_2 z + \varphi\right) \big],
\end{equation}
with $V$ the potential amplitude, $k_i=\pi/d_i (i=1,2)$ the lattice wave vectors, and $d_i$ the lattice periods.
The recoil energy of each lattice is $\Eri={\hbar^2 k_i^2}/{2m}$.
By controlling the ratio between the two wave vectors, we can realize either a periodic ($k_1=k_2, \varphi=0$) or quasiperiodic ($k_2/k_1=r$ with $r$ irrational, arbitrary $\varphi$) lattice. Here we use $r \simeq 1.2386...$.
In the experiment, the interaction strength, measured in terms of the Lieb-Liniger parameter $\gamma=m \gOneD/\hbar^2 n$, is controlled using Feshbach resonance techniques.

Figure~\ref{fig:main}(b) summarizes our main experimental and QMC results (detailed below).
It shows the quantum phase diagram of the Lieb-Liniger gas in the presence of either a pure periodic (commensurate case) or quasiperiodic potential (incommensurate case) with same total amplitude, at particle filling $nd_1=1$ ($nd_2=1/r<1$). The two axes indicate the total potential amplitude $V$ and the Lieb-Liniger parameter
$\gamma$.
We focus on the shallow lattice case where $V \sim \ErOne$.
For weak interaction and small lattice amplitude, both the periodic and quasiperiodic systems are superfluid (red region, SF).
For strong interaction and larger lattice amplitude, we find that both stabilize a MI with unit filling with respect to the first lattice (blue region).
These two domains are separated by a genuine phase transition but the
critical interaction strength $\gamma_{\textrm{c}}$ differs in the periodic and quasiperiodic cases.
The green region shows the domain where the periodic potential stabilizes a MI phase while the quasiperiodic potential induces a SF phase.
For the periodic case, the experimental data for the Mott transition (black disks) is the one of Ref.~\cite{Boeris2016}.
The demonstration of the Mott transition in the quasiperiodic case, \textit{i.e.}~the second transition line (corresponding to the open blue markers) between the green and blue regions is the first main result of our work. It is the first experimental demonstration of the Mott transition in a non-purely periodic atomic system.
The second one is the existence of the green region where the system
is a MI phase in the commensurate case and incommensurability restores a stable SF phase,
hence emphazing the crucial role of commensurability versus incommensurability.
This confirms the conjectured picture suggested in Fig.~\ref{fig:main}(a):
It indicates that the quasiperiodic lattice favors delocalization and induces a SF phase when the particle filling is commensurate with one of the lattice periods.
This is due to the fact that the presence of the second lattice blurs the periodicity of the first one, which is commensurate with the particle filling, since $nd_1=1$ but $nd_2<1$.
Note that the experimental and QMC results are consistent within errorbars.
All measurements shown in the manuscript have been realized by tuning the parameters so as to have a mean atomic density $\bar{n} \simeq 1/d_1$.
This way, even in the presence of harmonic trapping, the physics is controlled by the areas with density commensurate with the larger lattice spacing (see below).

\lettersection{Experimental measurement of the transition} 
The experiment starts with a Bose-Einstein condensate (BEC) with about $35 \times 10^3$ atoms of $^{39}$K with tunable scattering length~\cite{Roati2007}. It is then loaded into a strong 2D horizontal optical lattice, which splits the sample into about 1500 vertical tubes with a radial trapping frequency $\omega_{\perp}=2\pi \times 40$\,kHz. Each tube contains on average 33 atoms and has a longitudinal harmonic trap potential with frequency $\omega_z= 2\pi \times 160$\,Hz. Both inter and intra tubes, the atom distribution is inhomogeneous. The mean density $\bar{n}$ is then calculated by averaging overall the tubes \cite{DErrico2014}. Note our 3D BEC is prepared at temperature presumably lower than $10$\,nK. Suggested by recent works~\cite{yao-crossoverD-2023,guo-crossover-2023,guo-cooling-2023}, a further cooling may appear during the dimensional reduction.

We then adiabatically raise two weak vertical optical lattices with same amplitude ($V/2$) and different wavelengths $\lambda_{1}= 1064$\,nm  and $\lambda_{2}=859$\,nm to transfer the system in the shallow potential of Eq.~(\ref{eq:QPpotential}) with $k_i=2 \pi /\lambda_i$.
For the commensurate case, we set $\lambda_2=\lambda_1$.
We use $a_{load}=226 a_0$ ($a_0$ is the Bohr radius) to obtain a mean density $\bar{n} = (0.99 \pm 0.12)/d_1 = (0.80 \pm 0.09)/d_2$.
At this point we tune the scattering length to a variable value $a$ (hence varying $\gamma$) and explore the transport properties of the system in the $\gamma-V$ diagram.

By suddenly switching off a levitating vertical magnetic field gradient, we shift the center of the harmonic trap by $\delta z \approx$\,3\,$\mu$m and excite a sloshing motion of the system in the longitudinal (vertical) direction. After a variable evolution time $t$, we switch off all external confinements and let the atoms free to expand for $t_\textrm{\tiny TOF}=16.5$\,ms before time-of-flight (TOF) absorption images are recorded.

\begin{figure}[t] 
        \centering \includegraphics[width=1\columnwidth]{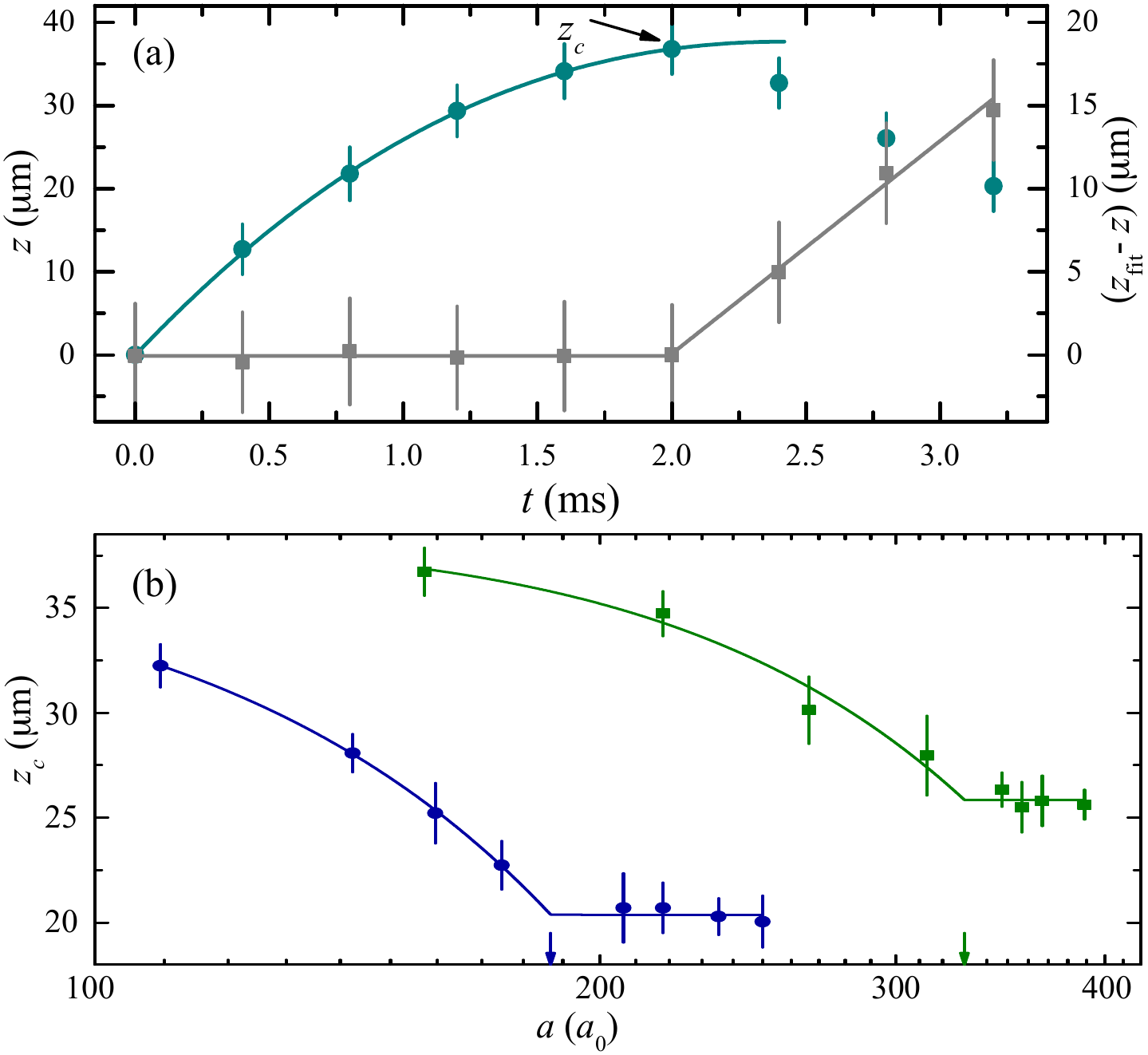}
        \caption{\label{fig:Exp} (a)~Evolution of the atomic peak position $z$ for a quasiperiodic potential with $V=(1.60 \pm 0.16)\,\ErOne$ and $a=(157 \pm 2)\,a_0$ (cyan circles, left axis) and difference between the fit of the evolution for $z<z_c$ ($z_{\mathrm{\tiny{fit}}}$) and the experimental data (gray squares, right axis). The error bars include a standard deviation of 4-6 independent measurements and the imaging resolution. The solid cyan (gray) line is a damped oscillation (piecewise) fit to the experimental data at short (at any) time. (b)~Critical value $z_c$ versus $a$ for two values of the potential depth: $V = (3.0 \pm 0.3) \ErOne$ (blue circles) and $V = (1.60 \pm 0.16) \ErOne$ (green squares). The error bars represent the statistical uncertainties. Solid lines are the piecewise fit to extract the critical value of the SF-MI critical point $a_c$ (arrows) and its uncertainty.
        }
\end{figure}

In Fig.~\ref{fig:Exp}(a) we show an example of the evolution of the atomic density peak position $z$ for fixed values of both the quasiperiodic potential depth $V$ and the scattering length $a$ (cyan circles). Typically we observe an initial increase of the peak position $z$ up to a certain critical value $z_c$, followed by a decrease for larger times due to phase-slip induced dissipation~\cite{Tanzi2016,Scaffidi2017,DErrico2017}. The data for $z<z_c$ are fitted by a damped oscillation $z_{\mathrm{\tiny{fit}}}(t)=z_\mathrm{\tiny{max}}e^{-Gt}\mathrm{sin}(\omega{'} t)$, with $\omega{'} = \sqrt{\omega_z^2m/m^*-G^2}$, $m^*$ the effective mass in the quasiperiodic potential, and $z_\mathrm{\tiny{max}}=\frac{m^* \omega^{*2} \delta z  t_\mathrm{\tiny TOF}}{m \omega^{'}}$. For larger times, the system enters a strongly dissipative regime where the phase slips nucleation rate diverges. By fitting the difference between the damped oscillation $z_{\mathrm{\tiny{fit}}}(t)$ and the experimental data at any time with a piecewise function (gray squares), we extract the critical value $z_c$ and its uncertainty. 
In the presence of a single periodic potential, the critical value $z_c$ is known to decrease with increasing interactions in both deep \cite{haller2010,Tanzi2013} and shallow \cite{Boeris2016,Scaffidi2017} optical lattices, and to vanish (reach a constant value) at the SF-MI transition in the former (latter) regime. We observe a similar behavior with the quasiperiodic lattice. 
In Fig.~\ref{fig:Exp}(b) we show the value $z_c$ for increasing scattering length $a$ for two values of the quasiperiodic potential depth $V$. As already observed for a single weak lattice potential \cite{Boeris2016}, $z_c$ decreases for increasing $a$ and reaches a plateau for $a>a_c$. The quantity $a_c$ can then be interpreted as the critical scattering length to enter the Mott lobe associated to the largest density ($1/d_1$).
We use a piecewise function with a second-order polynomial fit to extract $a_c$ and its uncertainty. Consistently with theoretical predictions, we find that $a_c$ increases as $V$ decreases.

In the case of a shallow quasiperiodic potential, theory predicts the existence of three different phases: the SF, the gapped MI, and the gapless BG. Nevertheless, the BG phase should appear only for potential depth $V$ larger than the critical value $V_c \approx 4.2 \ErOne$~\cite{Sanchez-Palencia2019,Sanchez-Palencia2020}, which is beyond our measured regime.
Hence, the observed fluid-insulator transition, being for $V<V_c$, suggests that the measured plateau corresponds to a MI. 

In Fig.~\ref{fig:main}(b) we show the critical interaction strength, expressed in terms of $\gamma$, for the superfluid-insulator transitions measured for several values of the potential depth $V$, in both periodic (black disks) and quasiperiodic (blue circles) cases.
In both cases, we find that $\gamma_c$ monotonically decreases with $V$ but the critical value to enter the MI regime is significantly higher for the incommensurate case than for the commensurate case, as discussed above.

\lettersection{Quantum Monte Carlo calculations}
We now discuss the numerical calculations.
Using continuous-space path integral quantum Monte Carlo~\cite{ceperley-PIMC-1995} with the worm algorithm implementation~\cite{boninsegni-worm-short-2006,boninsegni-worm-long-2006}, we simulate strongly-interacting 1D bosons in the presence of a shallow potential, be it periodic or quasiperiodic.
For given values of the chemical potential $\mu$, the temperature $T$, and the interaction strength $\gOneD$, we compute the particle density $n$ and superfluid fraction $\fs$.
The SF phase is characterized by $\fs>0$, whereas the MI phase corresponds to $\fs=0$ and a plateau at $nd_1=1$.
We use the system size $L=100 d_1$ and temperature $T \sim 3$\,nK. In practice, we judge the MI phase using the criteria $\fs<5\%$ and $|1-nd_1|<5\%$~\cite{Boeris2016}.
Note that we do not observe any cases where $\fs \simeq 0$ and $n d_1\neq 1$, consistently with the absence of a BG phase in the considered parameter range.

\begin{figure}[t!] 
          \centering \includegraphics[width=0.95\columnwidth]{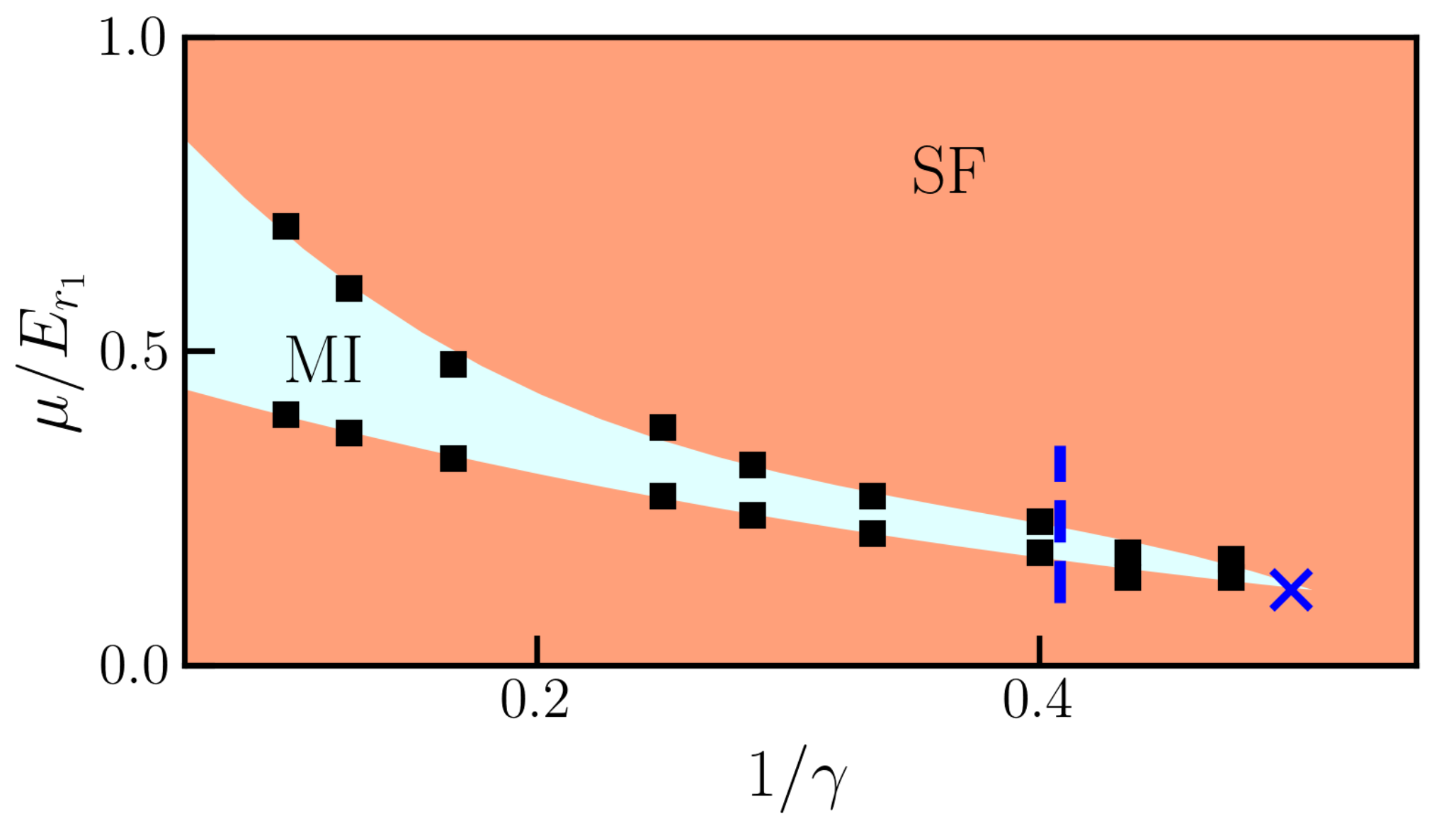}
           \caption{\label{fig:qmc}
           Phase diagram versus the inverse interaction parameter $1/\gamma$ and chemical potential $\mu$ for the quasiperiodic case with amplitude $V=3 \ErOne$, computed using QMC calculations.  The system size is $L=100 d_1$ and the temperature $T\sim3$\,nK. The MI lobe (blue), the SF region (red) and the critical point  $\gamma_\textrm{c}$ (blue cross), are determined from the SF-MI transition points (black squares). The dashed blue line indicates the coupling constant $\tilde{\gOneD}=2.46$ where the excitation gap is studied, see Fig.~\ref{fig:Spettri}.}
\end{figure}

In Fig.~\ref{fig:qmc}, we show the phase diagram versus inverse interaction strength $1/\gamma$ and chemical potential $\mu$ for the quasiperiodic case with $V=3 \ErOne$. We find a MI lobe with $nd_1=1$ (blue) surrounded by the SF phase (red). The SF-MI transitions are shown as black squares. At the tip of the Mott lobe, we use the resolution $\delta \mu=0.01\ErOne$ (such that $\delta \mu<\kB T$) and $\delta\gOneD=0.5\hbar^2/m$. This allows us to locate the critical value of $\gamma$  accurately. In this given example, we find $\gamma_{\textrm{c}}=2.0\pm 0.25$ (blue cross).

With this procedure, we compute the critical interaction strength $\gamma_{\textrm{c}}$ as a function of the potential amplitude $V$, for both periodic ($k_1=k_2,\ \varphi=0$) and quasiperiodic ($k_1\neq k_2, \varphi=0.2$) shallow potentials, see Eq.~(\ref{eq:QPpotential}). The final data are shown as solid black (quasiperiodic) and hallow blue (periodic) squares in Fig.~\ref{fig:main}(b).
We find that theory and experiment agree within error bars.
This further confirms the existence of a region (green) where the commensurate system
enters the MI phase, but incommensurability restores a stable SF.
The small deviation between theory and experiment may originate from the finite temperature and inhomogeneity effects in the experiment, and the finite resolution $\delta \mu$ in the numerics. Notably, here we observe a direct SF-MI transition up to our resolution. Whether there exists a BG sliver in between is an open question that deserves further analysis~\cite{vidal2001}.

\begin{figure}[t!] 
        \includegraphics[width=1\columnwidth]{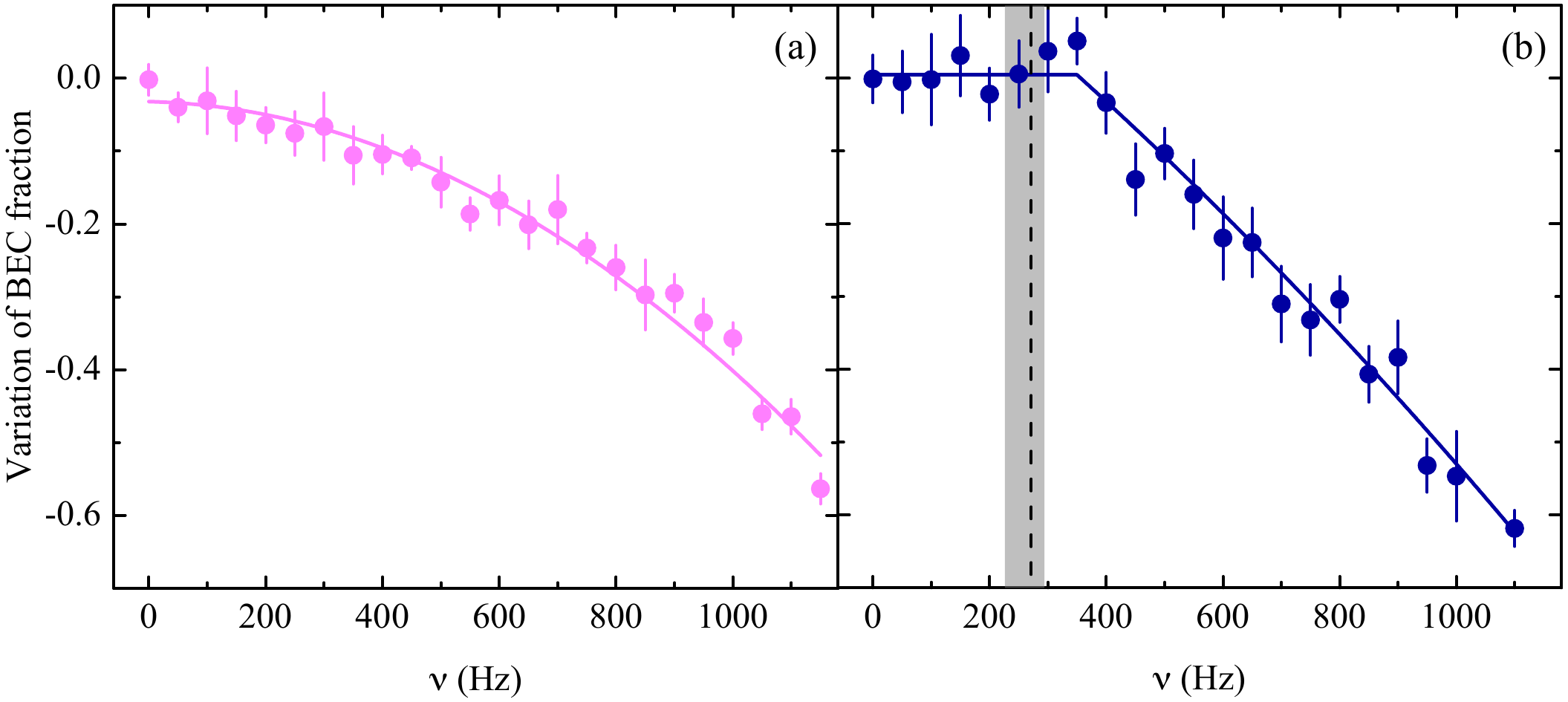}
        \caption{\label{fig:Spettri}
        Excitation spectra for a quasiperiodic lattice with depth $V=3\ErOne$ and two different scattering lengths (a)~below and (b)~above the critical value $a_c$ for the fluid-insulator transition: $a=(142.4 \pm 1.9)a_0$ and $a=(235 \pm 5)a_0$, respectively. 
        The error bars are a standard deviation of about 8 independent measurements. Solid lines are fits with a second order polynomial piecewise function. The vertical dashed line corresponds to the theoretical prediction for $a=235\ a_0$ and  $T=3$\,nK, while the gray area indicates its uncertainty, resulting from the systematic error on $a$ and $\ell_{\perp}$ in the experiment and the errorbar from QMC.
        }
\end{figure}

\lettersection{Detection of the Mott gap}
The MI phase is further characterized by the emergence of a finite gap in the excitation spectrum.
We modulate the depth of the quasiperiodic potential as $V (t)=V [1 + A \mathrm{cos}(2\pi\nu t)]$, with $A \simeq 0.1$ for $200$\,ms so as to generate excitations at the frequency $\nu$~\cite{DErrico2014}.
We then transfer back the Bose gas into the 3D optical trap by switching off all lattice beams and measure the variation of the BEC fraction as a function of the modulation frequency $\nu$~\cite{kollath2006,orso2009}.
Figure~\ref{fig:Spettri} shows two characteristic results respectively below and above the localization transition.
As expected, for weak interactions, any small modulation frequency is able to excite the system, corresponding to a gapless SF phase [Fig.~\ref{fig:Spettri}(a)].
In contrast, for strong interactions, no excitation is observed up to some frequency gap, corresponding to the MI phase [Fig.~\ref{fig:Spettri}(b)].
For $V=3 \ErOne$, the measured gap is $\nu_g=(350 \pm 50)$\,Hz and is consistent, within error bars, with the Mott gap calculated in the QMC simulations for the same parameters as in the experiment and low T [see dashed blue vertical line in Fig.~\ref{fig:qmc} and dashed black line in Fig.~\ref{fig:Spettri}(b)]. This further corroborates that the observed insulating phase is a gapped MI with density $n=1/d_1$. 

\lettersection{Discussion}
We have demonstrated the onset of a Mott transition in a shallow 1D quasiperiodic lattice. The experimental determination of the critical point as well as the Mott gap are in good quantitative agreement with ab initio QMC calculations.
In the range $1<V/\ErOne<3$ and for $\bar{n}=1/d_1$, the critical value of the interaction strength is larger in the incommensurate case compared to the commensurate case. It shows that in this regime incommensuration due to the quasiperiodic potential stabilizes the SF phase, i.e.~delocalization.
This may be qualitatively understood within a pertubative renormalization group picture: For $nd_1=1$, only the first lattice is relevant while the second lattice is irrelevant, within first order approximation, hence lowering the effective strength of the pinning potential. Still, the SF-to-MI transition observed in the incommensurate case is not quantitatively similar to that for a single lattice, which implies significant renormalization of the effective strength of the first lattice by the second. This mechanism may prefigure the onset of a BG phase for stronger potential amplitudes.

Our work provides an essential contribution to understanding the pivotal role played by commensuration in quantum phase transitions. Moreover, further experimental control of the temperature and flat-top potentials would allow to investigate more complex quantum phases, such as fractional MIs and BG phases~\cite{Sanchez-Palencia2019,Sanchez-Palencia2020,gautier-2Dquasicrystal-2021,zhu-2dquasicrystal-2022}.

\begin{acknowledgments}
This research was supported by the QuantERA Programme that has received funding from the European Union's Horizon 2020 research and innovation programme under Grant Agreements 731473 and 101017733, project MAQS, with funding organisation Consiglio Nazionale delle Ricerche; by the Agence Nationale de la Recherche (ANR, project ANR-CMAQ-002 France~2030), the program ``Investissements d'Avenir'' LabEx PALM (project ANR-10-LABX-0039-PALM), the Swiss National Science Foundation under grant numbers 200020-188687 and 200020-219400. HY, LSP and TG  would like to thank the Institut Henri Poincar\'e (UAR 839 CNRS-Sorbonne Universit\'e) and the LabEx CARMIN (ANR-10-LABX-59-01) for their support. Numerical calculations make use of the ALPS scheduler library and statistical analysis tools~\cite{troyer1998,ALPS2007,ALPS2011}.
\end{acknowledgments}

\end{document}